\documentstyle[seceq]{ptptex}

\title{
Cubic Matrices, Generalized Spin Algebra and Uncertainty Relation
}

\author{
Yoshiharu {\sc Kawamura}\footnote{E-mail:
haru@azusa.shinshu-u.ac.jp} %
}

\inst{
Department of Physics, Shinshu University, Matsumoto 390-8621, Japan
}

\recdate{April 16, 2003
}

\abst{
We propose a generalization of spin algebra using three-index objects.
Our results suggest the possibility that a triple commutation relation 
among three-index objects
implies a kind of uncertainty relation involving their expectation values.
}

\begin{document}

\maketitle

\section{Introduction}

Matrices play central roles in many branches of both mathematics and physics.
One of the main reasons for this is the following.
Physical systems are often described by many variables, some of which are
treated on an equal footing, e.g., spatial coordinates.
Systems often possess symmetries under certain transformations among such variables,
and such symmetry transformations, in many cases, form a group; e.g., 
rotations of spatial coordinates form a rotational group.
The use of matrices makes the analysis of systems with many variables simple and systematic,
because the action of such group elements can be represented by matrices.
Group theoretical analysis has been applied successfully to a wide range of systems, e.g.,
the classification of elementary particles
and the determination of the interactions among them.\cite{G}
Considering its success, it is natural to ask the following questions:
\begin{enumerate}
\item Is there a generalization of matrices?

\item If there is such a generalization, what are its advantages and applications 
for mathematics and physics?
\end{enumerate}

With regard to the first question, one might consider a many-index object, $A_{m_1 m_2 ...m_n}$,
since a matrix is a two-index object.
With regard to the second question, 
we point out that a new mechanics has been proposed that is
based on many-index objects\cite{YK},
and its basic structure has been studied from an algebraic point of view\cite{YK2,YK3}.
This mechanics has a counterpart in the canonical structure of classical mechanics
or Nambu mechanics\cite{Nambu},
and can be interpreted as its $\lq$quantum' or $\lq$discretized' version.
It can also be regarded as a generalization of Heisenberg matrix mechanics,
because a generalization of the Ritz rule is employed as a guiding principle.
This mechanics has very interesting properties, but
it is not yet clear whether it is applicable to real physical systems
and what physical meaning many-index objects possess.
It may also be possible to examine the algebraic structure 
of many-index objects independently of its dynamics to obtain information 
concerning their physical meaning and their role in describing physical phenomena.
This is the main motivation of our work.

In this paper, we propose a generalization of spin algebra using three-index objects and 
find that a certain triple commutation relation may lead to
a kind of uncertainty relation.

This paper is organized as follows.
In the next section, we give the definition of a three-index object.
We study a generalization of spin algebra in $\S$3. 
In $\S$4, we investigate the connection between a particular triple commutation relation 
among three-index objects
and an uncertainty involving their expectation values.
Section 5 is devoted to conclusions and discussion.

\section{Cubic matrix}

Here we state our definition of a cubic matrix\footnote{
Awata, Li, Minic and Yoneya introduced many-index objects
to construct a quantum version of the Nambu bracket.\cite{ALMY}
They refer to the three-index object among these as a $\lq$cubic matrix'.
We use the same terminology here,
although the definition of the triple product we use is different from theirs.}
and define related terminology.
A cubic matrix is an object with three indices written $A_{lmn}$, 
which is a generalization
of a usual matrix written analogously as $B_{mn}$.
We refer to a cubic matrix whose elements possess cyclic symmetry (i.e. , $A_{lmn} = A_{mnl} = A_{nlm}$)
as a cyclic cubic matrix.
We define the hermiticity of a cubic matrix by the relation
$A_{l'm'n'} = A_{lmn}^{*}$ for odd permutations among indices
and refer to a cubic matrix possessing hermiticity as a hermitian cubic matrix.
Here, the asterisk indicates complex conjugation.
A hermitian cubic matrix is a special type of cyclic cubic matrix, because it satisfies the relations
$A_{lmn} = A_{mln}^{*} = A_{mnl} = A_{nml}^{*} = A_{nlm} = A_{lnm}^{*}$.
We refer to the following form of a cubic matrix as a normal form or a normal cubic matrix:
\begin{eqnarray}
A_{lmn} = \delta_{lm} a_{mn} + \delta_{mn} a_{nl} + \delta_{nl} a_{lm}  .
\label{AN} 
\end{eqnarray}
A normal cubic matrix is a special type of cyclic cubic matrix.
The elements of a cubic matrix are treated as $c$-numbers throughout this paper.

We define the triple product of cubic matrices $A_{lmn}$, 
$B_{lmn}$ and $C_{lmn}$ by
\begin{eqnarray}
(ABC)_{lmn} \equiv
\sum_k A_{lmk} B_{lkn} C_{kmn} .
\label{cubicproduct}
\end{eqnarray}
The resultant three-index object, $(A B C)_{lmn}$,
does not necessarily possess cyclic symmetry, even if $A_{lmn}$, $B_{lmn}$ and $C_{lmn}$
are cyclic.
Note that this product is, in general, neither commutative nor associative; that is,
$(ABC)_{lmn} \neq (BAC)_{lmn}$ and
$(AB(CDE))_{lmn} \neq (A(BCD)E)_{lmn} \neq ((ABC)DE)_{lmn}$.
The triple-commutator we consider is defined by
\begin{eqnarray}
&~& [A, B, C]_{lmn} \equiv (ABC + BCA 
+ CAB  - BAC - ACB - CBA)_{lmn} .
\label{T-comm}
\end{eqnarray}
The corresponding triple-anticommutator is defined by
\begin{eqnarray}
&~& \{A, B, C\}_{lmn} \equiv (ABC + BCA 
+ CAB + BAC + ACB + CBA)_{lmn} .
\label{T-anticomm}
\end{eqnarray}
If $A_{lmn}$, $B_{lmn}$ and $C_{lmn}$ are hermitian,
then $i[A, B, C]_{lmn}$ and $\{A, B, C\}_{lmn}$ are also hermitian cubic matrices. 

Next, we define the product of two cubic matrices $A_{lmn}$ and $B_{lmn}$ by
\begin{eqnarray}
(AB)_{lm} \equiv
\sum_k A_{lmk} B_{klm} .
\label{product}
\end{eqnarray}
If $A_{lmn}$ and $B_{lmn}$
are hermitian cubic matrices,
the two-index object $(AB)_{lm}$
possess hermiticity, i.e., $(AB)_{lm} = (AB)^{*}_{ml}$.
If $A_{lmn}$ and $B_{lmn}$
are cyclic cubic matrices,
they commute with respect to this product, i.e.,
$(AB)_{lm} = (BA)_{lm}$.

\section{Generalized spin algebra}

\subsection{Spin algebra}

Here we review the spin algebra ${\it su}(2)$.
This algebra is defined by
\begin{eqnarray}
[J^a, J^b] = i \hbar \varepsilon^{abc} J^c ,
\label{spin-alg}
\end{eqnarray}
where the $J^a$ $(a = 1, 2, 3)$ are spin variables, $\hbar$ is the reduced Planck constant,
and $\varepsilon^{abc}$ is the Levi-Civita symbol.
Matrices in the adjoint representation are the $3 \times 3$ matrices given by
\begin{eqnarray}
(J^a)_{mn} = -i \hbar \varepsilon^{amn} ,
\label{adj}
\end{eqnarray}
where each of the indices $m$ and $n$ runs from 1 to 3.
Matrices in the spinor representation are the $2 \times 2$ matrices given by
\begin{eqnarray}
(J^a)_{mn} =  \frac{\hbar}{2} (\sigma^a)_{mn} ,
\label{spinor}
\end{eqnarray}
where the $\sigma^a$ are Pauli matrices, and each of the indices $m$ and $n$ runs from 1 to 2.

In general, matrices in the spin $j$ representation are the following $N \times N$ matrices:
\begin{eqnarray}
&~& (J^1)_{mn} = \frac{\hbar}{2} \Bigl(\sqrt{n(N-n)} \delta_{m n+1} 
+ \sqrt{m(N-m)} \delta_{m n-1}\Bigr) , \nonumber \\
&~& (J^2)_{mn} = \frac{\hbar}{2i} \Bigl(\sqrt{n(N-n)} \delta_{m n+1} 
- \sqrt{m(N-m)} \delta_{m n-1}\Bigr) , \nonumber \\
&~& (J^3)_{mn} = \frac{\hbar}{2} (2m-N-1) \delta_{mn} .
\label{N*N}
\end{eqnarray}
Here, each of the indices $m$ and $n$ runs from 1 to $N = 2j +1$.
The Casimir operator $\vec{J}^2$ is given by
\begin{eqnarray}
(\vec{J}^2)_{mn} \equiv (J^1)^2_{mn} + (J^2)^2_{mn} + (J^3)^2_{mn} 
 = \hbar^2 j (j + 1) \delta_{mn} .
\label{casimir}
\end{eqnarray}
The spinor representation matrices (\ref{spinor}) are obtained from (\ref{N*N}) 
by setting $j = \frac{1}{2}$, and
the adjoint representation matrices (\ref{adj}) are obtained from (\ref{N*N}) 
by setting $j = 1$,
after making a suitable unitary transformation.

\subsection{Generalization}

We now generalize the spin algebra defined by (\ref{spin-alg}) using hermitian cubic matrices.
In analogy to (\ref{adj}), we define the $4 \times 4 \times 4$ matrices 
we consider as follows:
\begin{eqnarray}
(J^a)_{lmn} = -i \hbar_C \varepsilon^{almn} , ~~
(K^a)_{lmn} = \hbar_C |\varepsilon^{almn}|  .
\label{adj-cubic}
\end{eqnarray}
Here, each of the indices $a$, $l$, $m$ and $n$ runs from 1 to 4,
and $\hbar_C$ is a new physical constant.
The matrices $(J^a)_{lmn}$ and $(K^a)_{lmn}$ form the algebra defined by
\begin{eqnarray}
&~& [J^a, J^b, J^c] = - i \hbar_C^2 \varepsilon^{abcd} K^d , 
~~~~ [J^a, J^b, K^c] = - i \hbar_C^2 \varepsilon^{abcd} J^d ,  \nonumber \\
&~& [J^a, K^b, K^c] =  i \hbar_C^2 \varepsilon^{abcd} K^d ,  
~~~~ [K^a, K^b, K^c] =  i \hbar_C^2 \varepsilon^{abcd} J^d ,
\label{cubic-spin-alg}
\end{eqnarray}
where the indices $l$, $m$ and $n$ are omitted.
There exists a subalgebra of the algebra defined by (\ref{cubic-spin-alg})
whose elements are $J^a$, $J^b$, $J^c$ and $K^d$ (or 
$K^a$, $K^b$, $K^c$ and $J^d$),
where $a$, $b$, $c$ and $d$ are all distinct.
For example, the elements $G^a = (J^1, J^2, J^3, K^4)$ form the algebra defined by
\begin{eqnarray}
&~& [G^a, G^b, G^c] = - i \hbar_C^2 \varepsilon^{abcd} G^d .
\label{cubic-spin-alg-G}
\end{eqnarray}
We refer to the algebra defined by (\ref{cubic-spin-alg-G}) as a $\lq$cubic spin algebra'.
The elements $G^a$ satisfy the so-called $\lq$fundamental indentity':
\begin{eqnarray}
&~& [[G^a, G^b, G^c], G^d, G^e] = [[G^a, G^d, G^e], G^b, G^c] \nonumber\\
&~& ~~~~~~~~~~~~ + [G^a, [G^b, G^d, G^e], G^c] + [G^a, G^b, [G^c, G^d, G^e]] .
\label{fund-id}
\end{eqnarray}

 As the counterparts of the spinor representation matrices (\ref{spinor}),
we define four kinds of hermitian $3 \times 3 \times 3$ matrices:
\begin{eqnarray}
&~& (S^1)_{lmn} \equiv \frac{\hbar_C}{\sqrt{2}} |\varepsilon_{lmn}| , 
~~~~ (S^2)_{lmn} \equiv \frac{\hbar_C}{i\sqrt{2}} \varepsilon_{lmn} , 
\nonumber \\
&~& (S^3)_{lmn} \equiv \frac{\hbar_C}{\sqrt{2}} \Bigl(\delta_{lm} \xi_{mn} + \delta_{mn} \xi_{nl}
 + \delta_{nl} \xi_{lm}\Bigr) ,
\nonumber \\
&~& (S^4)_{lmn} \equiv \frac{\hbar_C}{\sqrt{2}} \Bigl(\delta_{lm} \zeta_{mn} 
+ \delta_{mn} \zeta_{nl} + \delta_{nl} \zeta_{lm}\Bigr) .
\label{S3*3*3}
\end{eqnarray} 
Here, each of the indices $l$, $m$ and $n$ runs from 1 to 3, 
$\xi_{mn} \equiv (\delta_{m1} - \delta_{m2}) \delta_{n3}$ and  
$\zeta_{mn} \equiv \delta_{m1} \delta_{n2} +  \delta_{m2} \delta_{n1}$.\footnote{
This choice is not unique, but one instance is
$\xi_{mn} = \epsilon_m \varepsilon_{12n}$
and $\zeta_{mn} = \epsilon_m \varepsilon_{mn3}$,  
where $\epsilon_m$ takes either the value 1 or $-1$.}
It is shown that the variables $(S^a)_{lmn}$ 
form the cubic spin algebra (\ref{cubic-spin-alg-G}).

As an example of $(N+1) \times (N+1) \times (N+1)$ matrices ($N \geq 3$) that form the cubic spin algebra
defined by (\ref{cubic-spin-alg-G}), we have
\begin{eqnarray}
&~& (G^1)_{lmn} = \frac{\hbar_C}{8^{1/4}} \Bigl( (\delta_{lm-1} + \delta_{lm+1} \Bigr.
+ \delta_{lm-N+1} + \delta_{lm+N-1}) \nonumber \\
&~& ~~~~~~~~~~~~~~~~~~~~~~~~~~~~~~~~~~~~~~~~ \cdot (1-\delta_{lN+1})(1-\delta_{mN+1})\delta_{nN+1}  \nonumber \\
&~& ~~~~~~~~~ + (\delta_{mn-1} + \delta_{mn+1}
+ \delta_{mn-N+1} + \delta_{mn+N-1})(1-\delta_{mN+1})(1-\delta_{nN+1})\delta_{lN+1} \nonumber \\
&~& ~~~~~~~~~ + (\delta_{nl-1} + \delta_{nl+1}
\Bigl. + \delta_{nl-N+1} + \delta_{nl+N-1})(1-\delta_{nN+1})(1-\delta_{lN+1})\delta_{mN+1} \Bigr) , \nonumber \\
&~& (G^2)_{lmn} = \frac{\hbar_C}{8^{1/4}i} \Bigl( (\delta_{lm-1} - \delta_{lm+1} \Bigr.
+ \delta_{lm-N+1} - \delta_{lm+N-1}) \nonumber \\
&~& ~~~~~~~~~~~~~~~~~~~~~~~~~~~~~~~~~~~~~~~~ \cdot (1-\delta_{lN+1})(1-\delta_{mN+1})\delta_{nN+1}  \nonumber \\
&~& ~~~~~~~~~ + (\delta_{mn-1} - \delta_{mn+1}
+ \delta_{mn-N+1} - \delta_{mn+N-1})(1-\delta_{mN+1})(1-\delta_{nN+1})\delta_{lN+1} \nonumber \\
&~& ~~~~~~~~~ + (\delta_{nl-1} - \delta_{nl+1}
\Bigl. + \delta_{nl-N+1} - \delta_{nl+N-1})(1-\delta_{nN+1})(1-\delta_{lN+1})\delta_{mN+1} \Bigr) , \nonumber \\
&~& (G^3)_{lmn} = \frac{\hbar_C}{2^{1/4}} \Bigl( \delta_{lm} g^3_{mn}
+ \delta_{mn} g^3_{nl} + \delta_{nl} g^3_{lm} \Bigr) , \nonumber \\
&~& (G^4)_{lmn} = \frac{\hbar_C}{8^{1/4}} \Bigl( \delta_{lm} g^4_{mn}
+ \delta_{mn} g^4_{nl} + \delta_{nl} g^4_{lm} \Bigr) ,
\label{(N+1)*(N+1)*(N+1)}
\end{eqnarray}
where each of the indices $l$, $m$ and $n$ runs from 1 to $N = 2j +1$, and
$g^3_{mn}$ and $g^4_{mn}$ are defined by
\begin{eqnarray}
g^3_{mn} \equiv \epsilon_m (1 - \delta_{m N+1}) \delta_{n N+1} ~~~~~~~~~~~~~~~~~~~~~~~~~~~
\label{g3}
\end{eqnarray}
and
\begin{eqnarray}
&~& g^4_{mn} \equiv \epsilon_m (\delta_{mn-1} - \delta_{mn+1} + \delta_{mn-N+1} - \delta_{mn+N-1}) \nonumber \\
&~&  ~~~~~~~~~~~~~~~~~~~~ \cdot (1-\delta_{mN+1})(1-\delta_{nN+1}) .
\label{g4}
\end{eqnarray}
Here $\epsilon_m$ takes either the value 1 or $-1$.

Before ending this section, we note that
there exist $N \times N$ matrices $(M^a)_{mn}$ that satisfy the relations
$[M^a, M^b, M^c]_{mn} = - i \hbar^2 \varepsilon^{abcd} (M^d)_{mn}$.
Here, the triple commutator $[M^a, M^b, M^c]_{mn}$ is defined by
\begin{eqnarray}
&~& [M^a, M^b, M^c]_{mn} \equiv (M^aM^bM^c + M^bM^cM^a + M^cM^aM^b \nonumber \\
&~& ~~~~~~~~~~~~~~~~~~~~~~~~~~~  - M^bM^aM^c - M^aM^cM^b - M^cM^bM^a)_{mn} \nonumber \\
&~& ~~~~~~~~~~~~~~~~~~~~~ =  ([M^a, M^b] M^c)_{mn} + ([M^b, M^c] M^a)_{mn} \nonumber \\
&~& ~~~~~~~~~~~~~~~~~~~~~~~~~~~~~~~~~~~~~ + ([M^c, M^a] M^b)_{mn} ,
\label{T-comm-matrix}
\end{eqnarray}
with the usual definition of the triple product of matrices,
\begin{eqnarray}
(M^aM^bM^c)_{mn} \equiv \sum_{k,l} (M^a)_{mk} (M^b)_{kl} (M^c)_{ln} . 
\label{T-pro-matrix}
\end{eqnarray}
As an example, we have
\begin{eqnarray}
&~& (M^a)_{mn} = \frac{1}{(j(j+1))^{1/4}} (J^a)_{mn} ~~~~~ (a = 1 - 3), \nonumber \\
&~& (M^4)_{mn} = - (j(j+1))^{1/4} \hbar \delta_{mn} ,
\label{Na}
\end{eqnarray}
where $(J^a)_{mn}$ is given in (\ref{N*N}).

\section{Uncertainty relation}

\subsection{Uncertainty relation in quantum mechanics}

The uncertainty relation in quantum mechanics is expressed by
\begin{eqnarray}
\delta x \delta p \geq \frac{\hbar}{2} ,
\label{H-uncertainty}
\end{eqnarray}
where $\delta x$ and $\delta p$ represent uncertainties in the position 
and its canonical momentum, respectively.
This relation and a generalization can be elegantly formulated
in the present framework, as we now demonstrate.\cite{Uncertainty}

For any observable $A$, there is a corresponding hermitian operator
$\hat{A}$.
The expectation value of $A$ is defined by
\begin{eqnarray}
\langle A \rangle \equiv \int \psi^*(\vec{r}) A \psi(\vec{r}) d^3r , 
\label{<A>}
\end{eqnarray}
where $\psi(\vec{r})$ is a wave function that describes the state of the system.
The uncertainty in the value of a measurement for $A$ is defined by
\begin{eqnarray}
\delta A  \equiv \sqrt{\langle (A - \langle A \rangle)^2 \rangle} 
= \sqrt{\langle A^2 \rangle - \langle A \rangle^2} .
\label{deltaA}
\end{eqnarray}
Here $\delta A$ is a standard deviation that 
represents the magnitude of the fluctuation about the mean value.
For any observables $A$ and $B$, the following uncertainty relation holds:
\begin{eqnarray}
\delta A  \delta B \geq \frac{1}{2} |\langle [A, B] \rangle| .
\label{H-uncertainty2}
\end{eqnarray}

Let us derive the above relation in the matrix formalism for later convenience.
The following relationship exists between
the matrix $A_{mn}$ in Heisenberg matrix mechanics 
and the hermitian operator $\hat{A}$:
\begin{eqnarray}
A_{mn} = \int \phi_m^*(\vec{r}) \hat{A} \phi_n(\vec{r}) d^3r .
\label{Amn}
\end{eqnarray}
Here, the function $\phi_n(\vec{r})$ constitute a complete set of orthonormal functions.\footnote{
Here, we treat the case of a discrete spectrum for simplicity, but
it is straightforward to extend the present argument to the case of a continuous spectrum.} From 
(\ref{<A>}) and (\ref{Amn}), the expectation value $\langle A \rangle$ is written
\begin{eqnarray}
\langle A \rangle = \sum_{m,n} a_m^* a_n A_{mn}  
\label{<A>-matrix}
\end{eqnarray}
for the wave function $\psi(\vec{r}) = \sum_{n} a_n \phi_n(\vec{r})$. 
In the same way, the expectation value of $\hat{A}\hat{B}$ is written
\begin{eqnarray}
\langle A B \rangle  = \sum_{m,n}\sum_{k} a_m^* a_n A_{mk} B_{kn} \equiv (\vec{\cal A}, \vec{\cal B}) .
\label{<AB>}
\end{eqnarray}
Here, $\vec{\cal B}$ represents the complex vector whose $k$th component is $\sum_{n} B_{kn} a_n$.
Then, the uncertainty relation (\ref{H-uncertainty2}) can be demonstrated by use of
the Schwarz inequality $|\vec{\cal A}|^2 |\vec{\cal B}|^2 \geq |(\vec{\cal A}, \vec{\cal B})|^2$
and the relation $\hat{A}\hat{B} = \frac{1}{2} [\hat{A}, \hat{B}] + \frac{1}{2} \{\hat{A}, \hat{B}\}$.
Hence, this uncertainty relation is understood as a consequence of the
algebraic relation between physical variables.
If the expectation value $\langle [A, B] \rangle$ does not vanish,
it is not possible to measure the values of $A$ and $B$ simultaneously.
The uncertainty relation (\ref{H-uncertainty}) can be derived
from the commutation relation 
$[\hat{x}, \hat{p}] = i\hbar$.

\subsection{Generalized uncertainty relation}

We have seen that the uncertainty relation 
$\delta A  \delta B \geq \frac{1}{2} |\langle C \rangle|$
can be derived from the commutation relation $[A, B]_{mn} = i C_{mn}$ in quantum mechanics.
Then, it is natural to ask whether there is a similar uncertainty relation originating in 
a triple commutation relation $[A, B, C]_{lmn} = i D_{lmn}$.
(A typical such triple commutation relation is the cubic spin algebra discussed in the previous section.)
In the following, we find that an inequality of the form
$\delta A  \delta B \delta C \geq \frac{1}{6} |\langle D \rangle|$
indeed can be derived for certain types of definitions of the expectation values of many-index objects.

We define the expectation value of a cubic matrix $A_{lmn}$ by
\begin{eqnarray}
\langle A \rangle_c \equiv  \sum_{l,m,n} |a_l a_m a_n| 
e^{i(\theta_{lm} + \theta_{mn} + \theta_{nl})} A_{lmn} , 
\label{cubic<A>}
\end{eqnarray}
where $a_l$ is a complex number and
$\theta_{lm}$ is a real antisymmetric object: $\theta_{lm} = - \theta_{ml}$.
Then, the expectation value of $(A B C)_{lmn}$ is given by
\begin{eqnarray}
&~& \langle A B C \rangle_c  = \sum_{l,m,n} |a_l a_m a_n| 
e^{i(\theta_{lm} + \theta_{mn} + \theta_{nl})} \sum_{k} A_{lmk} B_{lkn} C_{kmn} \nonumber \\
&~& ~~~~~~~~~~~ \equiv \sum_{l,m,n} \sum_k {\cal{A}}^{(k)}_{lm} {\cal{B}}^{(k)}_{nl} {\cal{C}}^{(k)}_{mn} 
= \sum_{k} {\rm Tr}({\cal{A}}^{(k)} {\cal{C}}^{(k)} {\cal{B}}^{(k)}), 
\label{<ABC>}
\end{eqnarray}
where ${\cal{A}}^{(k)}_{lm} \equiv |a_l a_m|^{1/2} e^{i \theta_{lm}} A_{lmk}$.
Further, we define the expectation value of a two-index object $B_{lm}$ by
\begin{eqnarray}
\langle B \rangle_s \equiv  \sum_{l, m} |a_l a_m| 
e^{i(\theta_{lm} - \theta_{ml})} B_{lm} . 
\label{square<B>}
\end{eqnarray}
The expectation value of $(A^2)_{lm}$ is given by
\begin{eqnarray}
&~& \langle A^2 \rangle_s  = \sum_{l,m} |a_l a_m|
e^{i(\theta_{lm} - \theta_{ml})} \sum_{k} A_{lmk} A_{klm} \nonumber \\
&~& ~~~~~~~ = \sum_{l,m}\sum_k {\cal{A}}^{(k)}_{lm} {\cal{A}}^{(k)}_{lm}
= \sum_{l,m}\sum_k {\cal{A}}^{(k)}_{lm} {\cal{A}}^{*(k)}_{ml} \equiv |{\cal{A}}^{(k)}|^2.  \label{<A2>}
\end{eqnarray}
In this way, two kinds of expectation values\footnote{
In the case that $\theta_{lm} = \frac{1}{2}(\beta_m - \beta_l)$,
the expectation values $\langle A \rangle_c$ and $\langle B \rangle_s$ are reduced to
$\langle A \rangle_c \equiv  \sum_{l,m,n} |a_l a_m a_n|  A_{lmn}$ and
$\langle B \rangle_s \equiv  \sum_{l,m} a_l^* a_m  B_{lm}$,
where $a_m = |a_m| e^{i\beta_m}$.}
have been obtained without any guiding principle.
It is important to elucidate their physical meanings.
Next, by use of the inequality
\begin{eqnarray}
\Big(\sum_{k_1} |{\cal{A}}^{(k_1)}|^2\Big) \Big(\sum_{k_2} |{\cal{B}}^{(k_2)}|^2\Big) 
\Big(\sum_{k_3} |{\cal{C}}^{(k_3)}|^2\Big)
\geq |\sum_{k} {\rm Tr}({\cal{A}}^{(k)} {\cal{C}}^{(k)} {\cal{B}}^{(k)})|^2
\label{G-Schwarz}
\end{eqnarray}
and the relation
\begin{eqnarray}
\langle A B C \rangle_c  = \frac{1}{6} 
(\langle [A, B, C] \rangle_c + \langle \{A, B, C\} \rangle_c) ,
\label{ABC-rel}
\end{eqnarray}
the uncertainty relation
\begin{eqnarray}
\delta A  \delta B \delta C \geq \frac{1}{6} |\langle [A, B, C] \rangle_c| 
= \frac{1}{6} |\langle D \rangle_c| 
\label{G-uncertainty}
\end{eqnarray}
is derived, where the uncertainty $\delta A$ is defined by
\begin{eqnarray}
\delta A  \equiv \sqrt{\langle (A - \langle A \Delta \rangle_s \Delta)^2 \rangle_s} .
\label{deltaA-cubic}
\end{eqnarray}
Here $\Delta_{lmn} = \delta_{lm} \delta_{mn}$, and 
hence $\langle A \Delta \rangle_s = \sum_{m} |a_m|^2 A_{mmm}$.
Note that there exists the identity $[A, B, \Delta]_{lmn} =0$
for arbitrary cyclic cubic matrices $A_{lmn}$ and $B_{lmn}$. 
Finally, we discuss the physical implication of the uncertainty relation
(\ref{G-uncertainty}).
Let us assume that the 4-dimensional space-time coordinates 
are described by cubic matrices $(X^{\mu})_{lmn}$ ($\mu = 0,1,2,3$) 
that satisfy the relation
\begin{eqnarray}
[X^1, X^2, X^3]_{lmn} = - i l_{P}^2 (X^0)_{lmn} ,
\label{space-time-T-comm}
\end{eqnarray}
where $l_P$ is the Planck length, defined by $l_P \equiv \sqrt{2 G \hbar/c^3}$.
Here $G$ is the Newton constant and $c$ is a speed of light. From
the above argument, the following uncertainty relation can be derived:
\begin{eqnarray}
\delta X^1 \delta X^2 \delta X^3 \geq \frac{l_P^2}{6} |\langle X^0 \rangle| .
\label{space-time-uncertainty}
\end{eqnarray}
Many people have studied uncertainty relations
concerning the measurement of space-time distances
on the basis of various kinds of thought experiments.\cite{Review,K,NvD,A-C,NS,Y}
Among them, relations like 
$(\delta r)^3$ \protect\raisebox{-0.5ex}{$\stackrel{\scriptstyle >}{\sim}$} 
$l_P^2 r \sim l_P^2 c \delta t$ derived in Refs. \citen{K,NvD,A-C,NS}
are deeply related to (\ref{space-time-uncertainty}).
Here, $\delta r$ and $\delta t$ are the uncertainty in a spatial distance $r$ and a time period
as seen by an observer.
The following Lorentz covariant form is proposed in Ref. \citen{NS}:
\begin{eqnarray}
|\varepsilon_{\mu\nu\rho\sigma} n^{\mu} \delta x_{i_1}^{\nu} \delta x_{i_2}^{\rho} \delta x_{i_3}^{\sigma}|~
\protect\raisebox{-0.5ex}{$\stackrel{\scriptstyle >}{\sim}$}~ l_P^2 \delta x_{i_4}^{\mu} n_{\mu} .
\label{space-time-uncertainty2}
\end{eqnarray}
The quantities $\delta x_{i}^{\mu}$ here are the four-vectors defining a space-time volume, and
$n_{\mu}$ is any four-vector that represents the velocity of an observer.
The space-time uncertainty relation
(\ref{space-time-uncertainty2}) can be regareded as originating
from an algebraic relation of the form
\begin{eqnarray}
[X^{\mu}, X^{\nu}, X^{\rho}]  = - i l_{P}^2 \varepsilon^{\mu\nu\rho\sigma} X_{\sigma} .
\label{X-rel}
\end{eqnarray}

\section{Conclusions and discussion}

We have studied the generalization of spin algebra using cubic matrices.
Our results suggest the possibility that a triple commutation relation among cubic matrices implies
a kind of uncertainty relation involving their expectation values.
As a physical implication, we hypothesize that the space-time uncertainty relation
is connected to a triple commtation relation
of the form $[X^1, X^2, X^3]_{lmn} = - i l_{P}^2 (X^0)_{lmn}$.

The physical meaning of cubic matrices is not yet completely understood, and
there exist several questions.
The matrices $(J^a)_{mn}$ are representation matrices that operate on 
a representation space called the spin space.
It is yet unclear whether matrices $(G^a)_{lmn}$ also act as generators and
what kind of representation space exists for them.
In quantum mechanics, a matrix element $A_{mn}$ is interpreted 
as a probability amplitude between the state 
described by $\phi_m$ and that described by $\phi_n$.
The physical meaning of a cubic matrix element $A_{lmn}$, however, is not yet known.
Further, the derivation of the uncertainty relation (\ref{G-uncertainty}) seems tricky,
because the definition of the expectation values appears to be ad hoc.
We need to clarify the meanings of $a_l$ and $\theta_{lm}$.
To elucidate such points, it may be most useful to consider closely the 
physical meaning of expectation values.
It is also important to determine the relationship 
between the space-time uncertainty relations derived
from string/M theories \cite{Y} and those given in (\ref{space-time-uncertainty}).

\section*{Acknowledgements}
We would like to thank Professor S. Odake for useful discussions.

\end{document}